\documentclass{iaus}
\usepackage{graphicx}
\usepackage{hyperref}

\title[Around Gaia Alerts in 20 questions] %% give here short title %%
{Around Gaia Alerts in 20 questions}

\author[{\L}. Wyrzykowski \& S.T. Hodgkin]   %% give here short author list %%
{{\L}ukasz Wyrzykowski$^{1,2}$\footnote[2]{name pronunciation: {\it Woo-cash Vi-zhi-kov-ski}}
 \and Simon Hodgkin$^1$}

\affiliation{$^1$Institute of Astronomy, University of Cambridge, Madingley Road, CB3 0HA Cambridge, UK \\
email: {\tt wyrzykow@ast.cam.ac.uk}, {\tt sth@ast.cam.ac.uk}\\
[\affilskip]
$^2$ Warsaw University Astronomical Observatory, Al. Ujazdowskie 4, 00-478 Warszawa, Poland\\
email: {\tt lw@astrouw.edu.pl}\\
}

\pubyear{2011}
\volume{285}  %% insert here IAU Symposium No.
\pagerange{xxx--xxx}
\date{19-23 September 2011}
\jname{New Horizons in Time Domain Astronomy}
\editors{R.E.M. Griffin, R.J. Hanisch \& R. Seaman}
\begin{document}

\maketitle

\begin{abstract}

  Gaia is a European Space Agency (ESA) astrometry space mission, and
  a successor to the ESA Hipparcos mission. Gaia's main goal is to
  collect high-precision astrometric data (i.e. positions, parallaxes,
  and proper motions) for the brightest 1 billion objects in the
  sky. These data, complemented with multi-band, multi-epoch
  photometric and spectroscopic data collected from the same observing
  platform, will allow astronomers to reconstruct the formation
  history, structure, and evolution of the Galaxy.

  Gaia will observe the whole sky for 5 years, providing a unique
  opportunity for the discovery of large numbers of transient and
  anomalous events, e.g. supernovae, novae and microlensing events, GRB
  afterglows, fallback supernovae, and other theoretical or unexpected
  phenomena. The Photometric Science Alerts team has been tasked with
  the early detection, classification and prompt release of anomalous
  sources in the Gaia data stream. In this paper, we discuss the
  challenges we face in preparing to use Gaia to search for transient
  phenomena at optical wavelengths.

\end{abstract}

\subsubsection*{\bf 1. Where, how and when?}

\noindent
Gaia will be launched from ESA/Kourou (French Guyana) onboard a
Soyuz-Fregat rocket in June 2013.  Deployment will be at the L2
Lagrange Point, with the first community release of alerts expected in
mid 2014 (internal verification will begin in early 2014). The mission
is scheduled to end in 2018--2019.

\subsubsection*{\bf 2. What telescopes will Gaia have?}

\noindent 
Gaia will be equipped with two 1.45x0.5m primary mirrors, forming two
fields of view separated by 106.5 degrees. The light from both mirrors
will be imaged onto a single focal plane. Gaia will reach down to V=20
in the Astrometric Field detectors.

\subsubsection*{\bf 3. What instruments will Gaia have?}
\begin{figure}[h]
% \vspace*{-2.0 cm}
\begin{center}
 \includegraphics[width=4.4in]{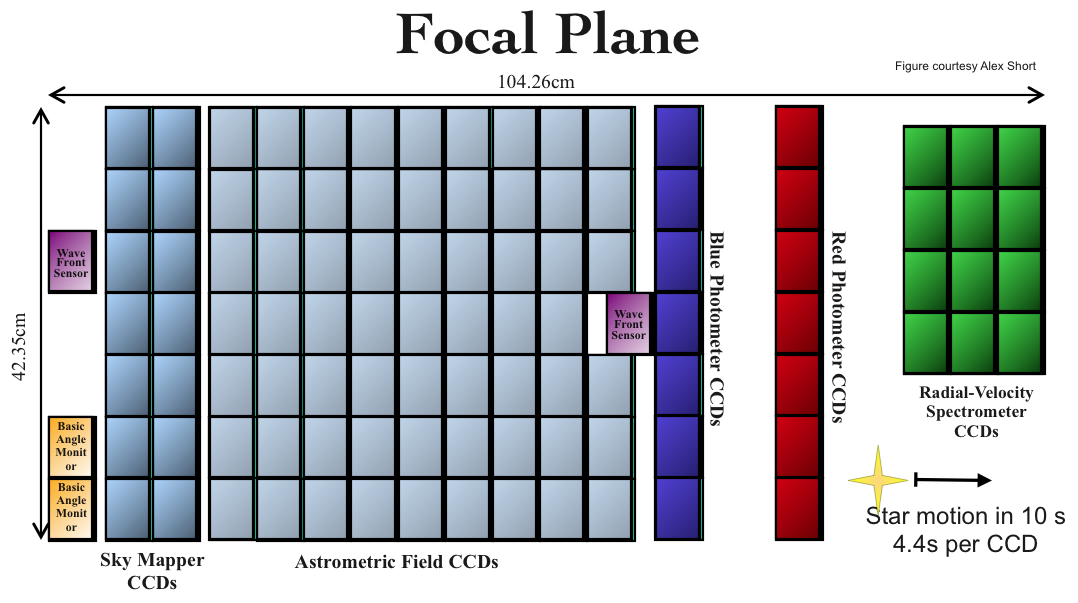}
% \vspace*{-1.0 cm}
 \caption{Focal plane of Gaia.}
   \label{fig:focalplane}
\end{center}
\end{figure}

\noindent 
Each object traverses through the focal plane (4.4 sec per CCD), see
Figure \ref{fig:focalplane}.

\noindent{\bf SM}: Objects are detected in Sky Mapper CCDs, and are
allocated windows for the remaining detectors.\\
{\bf AF}: Source positions and G-band magnitudes are measured in
the Astrometric Field CCDs (platescale $\sim 0.04 \times 0.1$
milliarcsecs).\\
{\bf BP/RP}: Low-dispersion spectro-photometry (330-680nm, 640-1000nm) in 120 samples.\\
{\bf RVS}: Intermediate-dispersion (R$\sim$11,500) spectroscopy
(847-874nm) around the Calcium Infrared triplet to V$<$17 mag.\\

\subsubsection*{\bf 4. What is the data latency?}

\noindent 
Gaia will be visible from the Earth for only 8h a day. All data from
the last 24h will be downlinked during a contact. After initial
processing, alerts will be issued from between a couple of hours, and
up to 48 hours, after the observation.

\subsubsection*{\bf 5. What is downloaded?}

\noindent 
Most of the sky is empty. Gaia will only transmit small windows around stars  detected at each transit on the Star Mapper CCDs  and associated data.  

\subsubsection*{\bf 6. How does the scanning law allow for full sky coverage?}

\begin{figure}[h]
\begin{center}
 \includegraphics[width=2.2in]{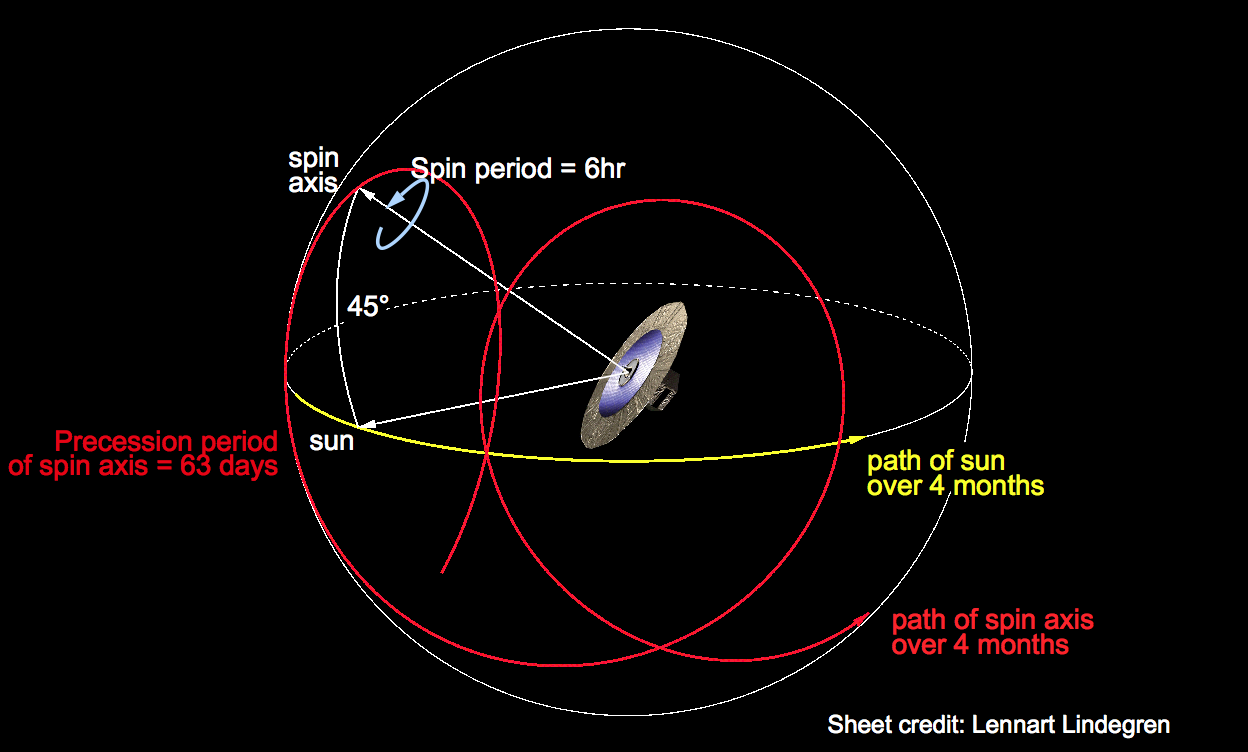} 
 \includegraphics[width=2.2in]{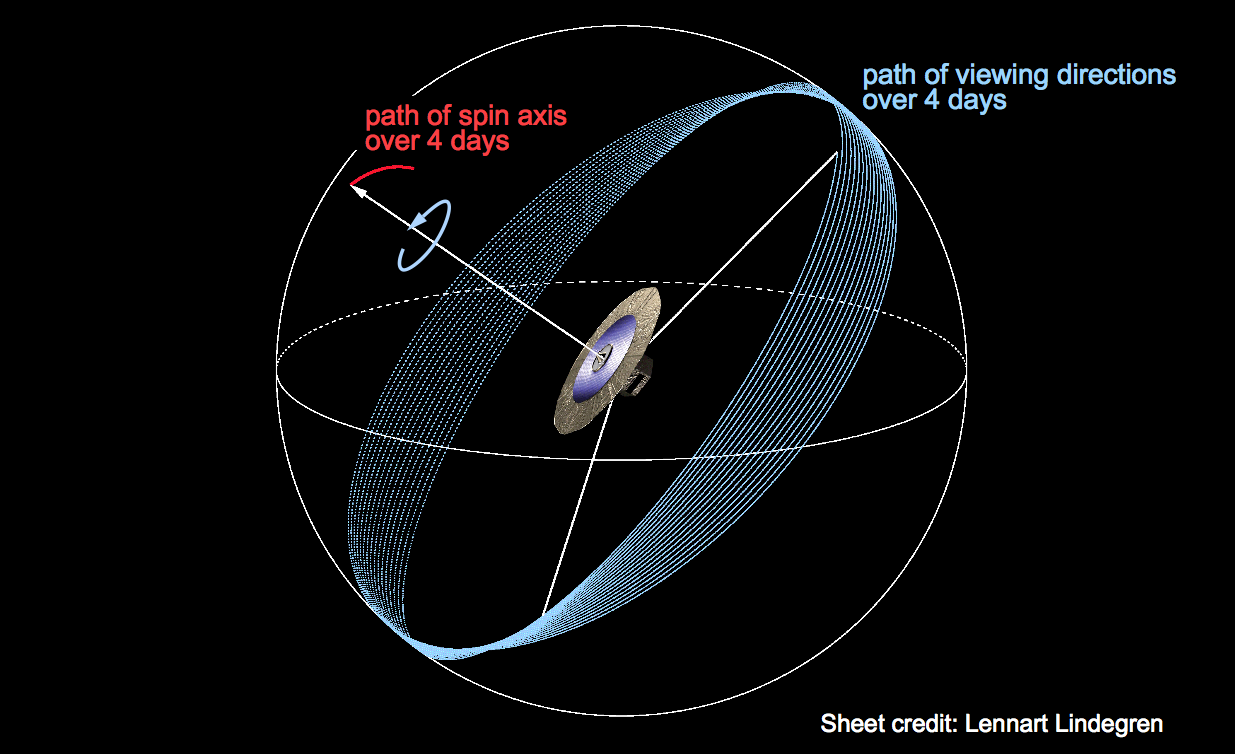} 
 \caption{Nominal Scanning Law principles for Gaia satellite.}
   \label{fig:scanninglaw}
\end{center}
\end{figure}

\noindent
Gaia has a pre-defined plan for scanning the sky. The spin axis is
maintained at a 45 deg angle from the Sun, with a period of 6h. For
details see Figure \ref{fig:scanninglaw}. 

\subsubsection*{\bf 7. What is the typical sampling?} 

\noindent 
On average, each object will be observed 80 times, though at the
Ecliptic nodes, objects are scanned in excess of 200
times. Observations occur in pairs (two FOVs), separated by $\sim$2
hours. The next pair will typically occur between 6 hours and $\sim$30
days later.

\subsubsection*{\bf 8. What is the precision of the instantaneous
  photometry and astrometry?}

\begin{figure}[h]
\begin{center}
 \includegraphics[width=2.6in]{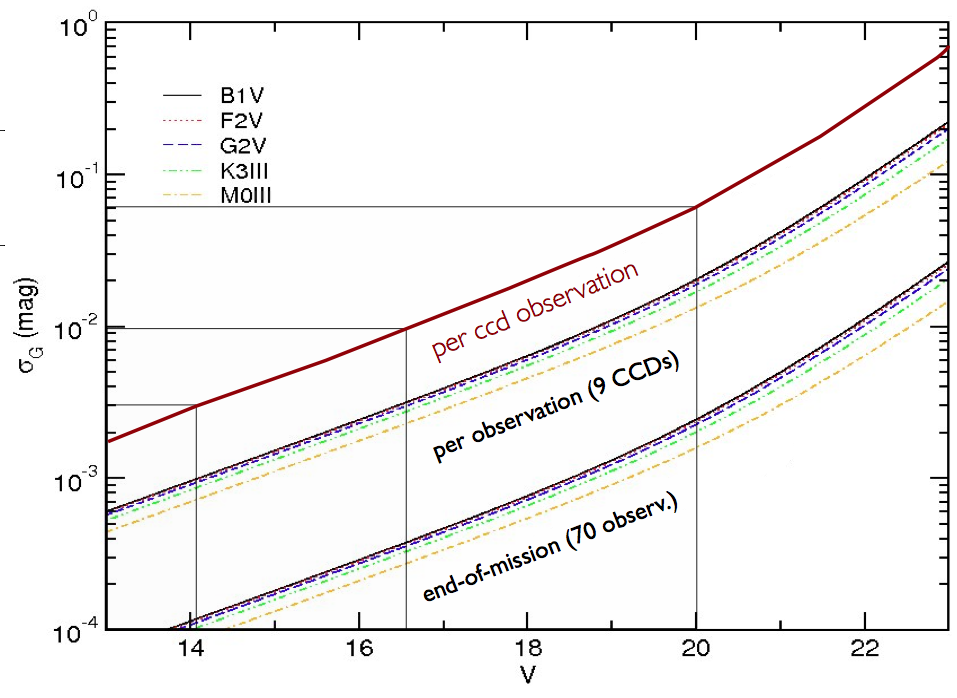}
 \includegraphics[width=2.6in]{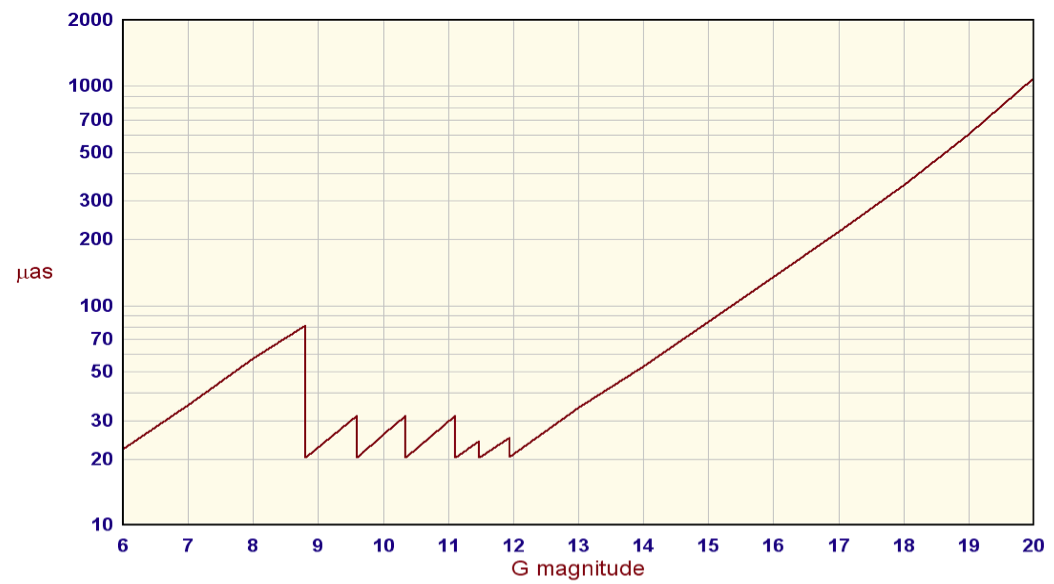}
 \caption{Precision of instantaneous photometry and astrometry of Gaia satellite. (from \cite{Varadi2009}).}
   \label{fig:precision}
\end{center}
\end{figure}

\noindent 
In a single observation (transit) the photometry will reach
milli-magnitude precision at G=14, and 1\% at G=19.  The astrometric
precision will be in the range 20-80$\mu$as at G=8-15 (see
Figure~\ref{fig:precision} for the effects of gating), falling to
600$\mu$as at G=19. This astrometric precision will only be reached
later in the mission.

\subsubsection*{\bf 9. How will the anomalies be detected?}

\noindent Using simple recipes:\\
1. Compare the most recent observation with the historic data
available.\\
2. Inspect for unexpected changes.\\
3. No history? - new transient!

\subsubsection*{\bf 10. How will the anomalies be classified?}

\noindent
1. From the light-curve.\\
2. Using low-dispersion BP/RP spectroscopy.\\
3. Cross-matching with archival data.

\subsubsection*{\bf 11. How will the BP/RP spectra be used?}

\noindent
Self-Organizing Maps (\cite{WyrzykowskiBelokurov2008}) built from the
low-dispersion spectra can confirm a non-stellar nature, classify
Supernova types, measure Supernova ages and possibly even constrain
the redshift.

\begin{figure}[h]
\begin{center}
 \includegraphics[width=3.5in]{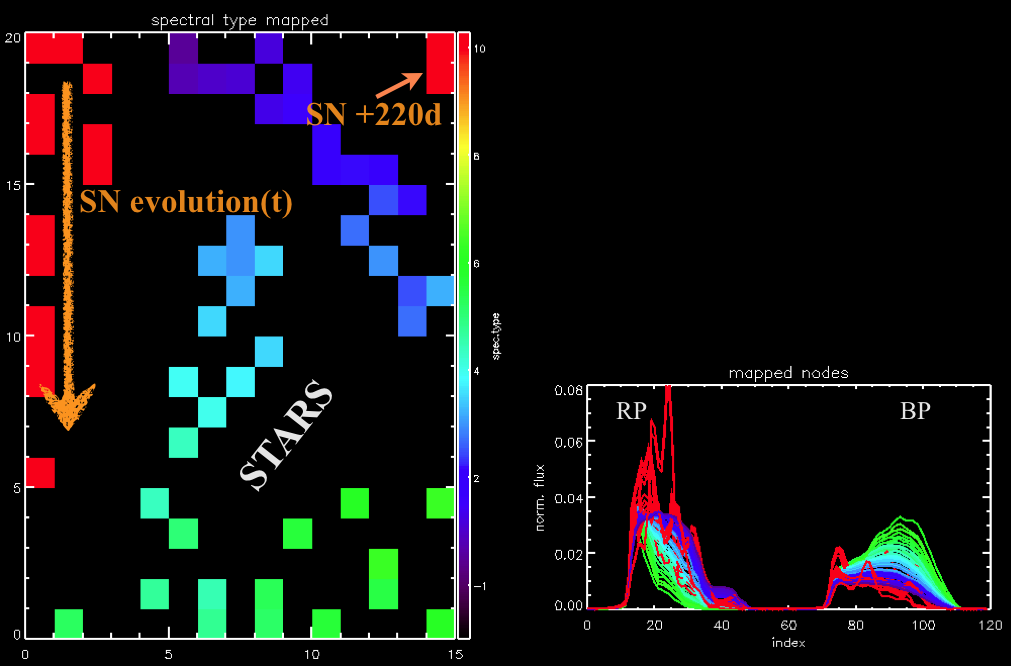}
 \caption{A Self-Organizing Map (left) can distinguish between different
   spectral types of stars and supernova at different epochs, as built
   from Gaia synthesized BP/RP spectra (right).}
   \label{fig:som}
\end{center}
\end{figure}

\subsubsection*{\bf 12. How will the alerts be disseminated?}

\noindent
Skyalert.org, email, www server, Twitter, iPhone app, etc.

\subsubsection*{\bf 13. What will be in an alert?}

\noindent
The coordinates, a small cutout image from the SM, the Gaia
light-curve, a low-resolution spectrum at the trigger, the
classification results, and the cross-matching results.

\subsubsection*{\bf 14. What will the the main triggers be?}

\noindent
Supernovae, Classical novae, dwarf novae, Microlensing events, Be
stars, GRB afterglows, M-dwarf flares, R CrB-type stars, FU Ori-type
stars, Asteroids, Suprises.

\subsubsection*{\bf 15. How many Supernovae will Gaia detect over 5 years?}

\noindent
6000 SNe expected down to G=19. About 2000 will be detected before the
maximum (\cite{BelokurovSN}).

\subsubsection*{\bf 16. How many Microlensing Events will Gaia detect?}

\noindent
1000+ events (mostly long $t_E>$30d) are expected to be detected
photometrically, mainly in the Galactic bulge and
plane. Astrometric centroid motion will be detectable
in real-time (for larger deviations of about 100$\mu$as) in on-going
events, and alerts may be triggered to obtain complementary photometry
(\cite{BelokurovMICROLENSING}).

\begin{figure}[h]
\begin{center}
 \includegraphics[width=4.3in]{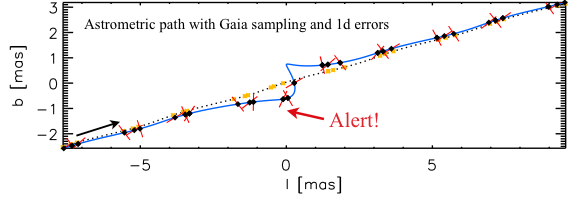} 
 \caption{Trajectory of a source due to proper motion and centroid shift during a microlensing event.}
   \label{fig:astrometricalert}
\end{center}
\end{figure}

\subsubsection*{\bf 17. Will Gaia alert on GRB optical counterparts?}

\noindent
Gaia sampling and data latency is not good for alerting on
GRBs. However, we still expect to detect 1-2 bright on-axis afterglows
and 5-15 orphan afterglows (\cite{Japelj2011}).

\subsubsection*{\bf 18. How many Asteroids will Gaia see?}

\noindent
About 250,000 asteroids (mostly known). Alerts on new asteroids and
NEO candidates will be based on unsuccessful star matching.

\subsubsection*{\bf 19. What about known anomalous objects?}

\noindent
Such objects can be added to the {\bf Watch List}. Every time Gaia
observes them, their data will become available for inspection.

\subsubsection*{\bf 20. How can I get involved now?}

\noindent
{\bf - with my telescope time:} prepare for Gaia Alerts,
register at Skyalert.org, set-up your alerts on CRTS stream
(\cite{DrakeCRTS}) (SNe, CVs, blazars, etc.), follow-up the alerts,
contact us with your data!

\noindent
 {\bf - with my scientific interests:} suggest what would be
worth detecting and alerting on, propose detection algorithms and
classification techniques, suggest interesting known targets to be
observed.

\subsubsection*{\bf More information on the web:}
\begin{itemize}
\item Gaia ESA web pages: \href{http://gaia.esa.int}{{\it http://gaia.esa.int}} \\and \href{http://www.rssd.esa.int/index.php?project=GAIA\&page=index}{{\it http://www.rssd.esa.int/index.php?project=GAIA\&page=index}}
\item Gaia Science Alerts Working Group wiki: \href{http://www.ast.cam.ac.uk/ioa/research/gsawg/}{{\it http://www.ast.cam.ac.uk/ioa/research/gsawg/}}
\item original poster on Gaia Alerts presented at the IAU Symposium in Oxford in September 2011: \href{http://www.ast.cam.ac.uk/ioa/wikis/gsawgwiki/index.php/Detection_system}{{\it http://www.ast.cam.ac.uk/ioa/wikis/gsawgwiki/index.php/Detection\_system}}
\end{itemize}

\subsubsection*{\bf Acknowledgement}
This work relies on efforts of numerous people involved in the preparations for the Gaia mission within Data Processing and Analysis Consortium (DPAC). Their work is acknowledged here and thanked for.


\begin{thebibliography}{}
\bibitem[Belokurov and Evans (2002)]{BelokurovSN}{Belokurov, V. A., 
Evans, N. W.} 2002,  \textit{MNRAS} 331,  649

\bibitem[Belokurov and Evans (2003)]{BelokurovMICROLENSING}{Belokurov, V. A., 
Evans, N. W.} 2003,  \textit{MNRAS} 341,  569 

\bibitem[Drake et al. (2009)]{DrakeCRTS}{Drake, A. J., 
Djorgovski, S. G., Mahabal, A., Beshore, E., Larson, S., Graham, M. J., 
Williams, R., Christensen, E., Catelan, M., Boattini, A., Gibbs, A., Hill, 
R., Kowalski, R.} 2009,  \textit{ApJ} 696,  870 

\bibitem[Japelj and Gomboc (2011)]{Japelj2011}{Japelj, J., Gomboc, A.} 2011,  \textit{PASP} 123,  1034

\bibitem[Varadi et al. 2009]{Varadi2009}{Varadi, M. Eyer, L., Jordan, S., Mowlavi, N., Koester, D.}2009, \textit{AIPC} 1170, 330, \href{http://arxiv.org/abs/0907.4084}{arXiv:0907.4084}

\bibitem[Wyrzykowski and Belokurov (2008)]{WyrzykowskiBelokurov2008}{Wyrzykowski, {\L}., Belokurov, V.} 2008,  
\textit{AIPC} 1082,  201, \href{http://arxiv.org/abs/0811.1808}{arXiv:0811.1808}

\end{thebibliography}
\end{document}